\documentclass[aps,prl,twocolumn,superscriptaddress]{revtex4-2}

\RequirePackage{graphicx}
\usepackage{multirow}
\usepackage{xcolor}
\usepackage{amsmath}
\usepackage{amssymb}
\usepackage{hyperref}
\usepackage{microtype} % To align text
\usepackage{booktabs}
\usepackage{multirow}
\usepackage{subcaption}
\usepackage{graphicx}
\usepackage{array} 
\newcolumntype{C}[1]{>{\centering\arraybackslash}p{#1}}

\begin{document}

%\preprint{}

\title{
Limits on WIMP dark matter with NaI(Tl) crystals in three years of COSINE-100 data
}


%\author{}
%\thanks{}
%\affiliation{}
\author{G.~H.~Yu}
\thanks{tksxk752@naver.com}
\affiliation{Department of Physics, Sungkyunkwan University, Suwon 16419, Republic of Korea}
\affiliation{Center for Underground Physics, Institute for Basic Science (IBS), Daejeon 34126, Republic of Korea}
\author{N.~Carlin}
\affiliation{Physics Institute, University of S\~{a}o Paulo, 05508-090, S\~{a}o Paulo, Brazil}
\author{J.~Y.~Cho}
\affiliation{Center for Underground Physics, Institute for Basic Science (IBS), Daejeon 34126, Republic of Korea}
\affiliation{Department of Physics, Kyungpook National University, Daegu 41566, Republic of Korea}
\author{J.~J.~Choi}
\affiliation{Department of Physics and Astronomy, Seoul National University, Seoul 08826, Republic of Korea} 
\affiliation{Center for Underground Physics, Institute for Basic Science (IBS), Daejeon 34126, Republic of Korea}
\author{S.~Choi}
\affiliation{Department of Physics and Astronomy, Seoul National University, Seoul 08826, Republic of Korea} 
\author{A.~C.~Ezeribe}
\affiliation{Department of Physics and Astronomy, University of Sheffield, Sheffield S3 7RH, United Kingdom}
\author{L.~E.~Fran{\c c}a}
\affiliation{Physics Institute, University of S\~{a}o Paulo, 05508-090, S\~{a}o Paulo, Brazil}
\author{C.~Ha}
\affiliation{Department of Physics, Chung-Ang University, Seoul 06973, Republic of Korea}
\author{I.~S.~Hahn}
\affiliation{Center for Exotic Nuclear Studies, Institute for Basic Science (IBS), Daejeon 34126, Republic of Korea}
\affiliation{Department of Science Education, Ewha Womans University, Seoul 03760, Republic of Korea} 
\affiliation{IBS School, University of Science and Technology (UST), Daejeon 34113, Republic of Korea}
\author{S.~J.~Hollick}
\affiliation{Department of Physics and Wright Laboratory, Yale University, New Haven, CT 06520, USA}
\author{E.~J.~Jeon}
\thanks{ejjeon@ibs.re.kr}
\affiliation{Center for Underground Physics, Institute for Basic Science (IBS), Daejeon 34126, Republic of Korea}
\affiliation{IBS School, University of Science and Technology (UST), Daejeon 34113, Republic of Korea}
\author{H.~W.~Joo}
\affiliation{Department of Physics and Astronomy, Seoul National University, Seoul 08826, Republic of Korea} 
\author{W.~G.~Kang}
\affiliation{Center for Underground Physics, Institute for Basic Science (IBS), Daejeon 34126, Republic of Korea}
\author{M.~Kauer}
\affiliation{Department of Physics and Wisconsin IceCube Particle Astrophysics Center, University of Wisconsin-Madison, Madison, WI 53706, USA}
\author{B.~H.~Kim}
\affiliation{Center for Underground Physics, Institute for Basic Science (IBS), Daejeon 34126, Republic of Korea}
\author{H.~J.~Kim}
\affiliation{Department of Physics, Kyungpook National University, Daegu 41566, Republic of Korea}
\author{J.~Kim}
\affiliation{Department of Physics, Chung-Ang University, Seoul 06973, Republic of Korea}
\author{K.~W.~Kim}
\affiliation{Center for Underground Physics, Institute for Basic Science (IBS), Daejeon 34126, Republic of Korea}
\author{S.~H.~Kim}
\affiliation{Center for Underground Physics, Institute for Basic Science (IBS), Daejeon 34126, Republic of Korea}
\author{S.~K.~Kim}
\affiliation{Department of Physics and Astronomy, Seoul National University, Seoul 08826, Republic of Korea}
\author{W.~K.~Kim}
\affiliation{IBS School, University of Science and Technology (UST), Daejeon 34113, Republic of Korea}
\affiliation{Center for Underground Physics, Institute for Basic Science (IBS), Daejeon 34126, Republic of Korea}
\author{Y.~D.~Kim}
\affiliation{Center for Underground Physics, Institute for Basic Science (IBS), Daejeon 34126, Republic of Korea}
\affiliation{IBS School, University of Science and Technology (UST), Daejeon 34113, Republic of Korea}
\author{Y.~H.~Kim}
\affiliation{Center for Underground Physics, Institute for Basic Science (IBS), Daejeon 34126, Republic of Korea}
\affiliation{IBS School, University of Science and Technology (UST), Daejeon 34113, Republic of Korea}
\author{Y.~J.~Ko}
\affiliation{Department of Physics, Jeju National University, Jeju 63243, Republic of Korea}
\author{D.~H.~Lee}
\affiliation{Department of Physics, Kyungpook National University, Daegu 41566, Republic of Korea}
\author{E.~K.~Lee}
\affiliation{Center for Underground Physics, Institute for Basic Science (IBS), Daejeon 34126, Republic of Korea}
\author{H.~Lee}
\affiliation{IBS School, University of Science and Technology (UST), Daejeon 34113, Republic of Korea}
\affiliation{Center for Underground Physics, Institute for Basic Science (IBS), Daejeon 34126, Republic of Korea}
\author{H.~S.~Lee}
\affiliation{Center for Underground Physics, Institute for Basic Science (IBS), Daejeon 34126, Republic of Korea}
\affiliation{IBS School, University of Science and Technology (UST), Daejeon 34113, Republic of Korea}
\author{H.~Y.~Lee}
\affiliation{Center for Exotic Nuclear Studies, Institute for Basic Science (IBS), Daejeon 34126, Republic of Korea}
\author{I.~S.~Lee}
\affiliation{Center for Underground Physics, Institute for Basic Science (IBS), Daejeon 34126, Republic of Korea}
\author{J.~Lee}
\affiliation{Center for Underground Physics, Institute for Basic Science (IBS), Daejeon 34126, Republic of Korea}
\author{J.~Y.~Lee}
\affiliation{Department of Physics, Kyungpook National University, Daegu 41566, Republic of Korea}
\author{M.~H.~Lee}
\affiliation{Center for Underground Physics, Institute for Basic Science (IBS), Daejeon 34126, Republic of Korea}
\affiliation{IBS School, University of Science and Technology (UST), Daejeon 34113, Republic of Korea}
\author{S.~H.~Lee}
\affiliation{IBS School, University of Science and Technology (UST), Daejeon 34113, Republic of Korea}
\affiliation{Center for Underground Physics, Institute for Basic Science (IBS), Daejeon 34126, Republic of Korea}
\author{S.~M.~Lee}
\affiliation{Department of Physics and Astronomy, Seoul National University, Seoul 08826, Republic of Korea} 
\author{Y.~J.~Lee}
\affiliation{Department of Physics, Chung-Ang University, Seoul 06973, Republic of Korea}
\author{D.~S.~Leonard}
\affiliation{Center for Underground Physics, Institute for Basic Science (IBS), Daejeon 34126, Republic of Korea}
\author{N.~T.~Luan}
\affiliation{Department of Physics, Kyungpook National University, Daegu 41566, Republic of Korea}
\author{V.~H.~A.~Machado}
\affiliation{Physics Institute, University of S\~{a}o Paulo, 05508-090, S\~{a}o Paulo, Brazil}
\author{B.~B.~Manzato}
\affiliation{Physics Institute, University of S\~{a}o Paulo, 05508-090, S\~{a}o Paulo, Brazil}
\author{R.~H.~Maruyama}
\affiliation{Department of Physics and Wright Laboratory, Yale University, New Haven, CT 06520, USA}
\author{R.~J.~Neal}
\affiliation{Department of Physics and Astronomy, University of Sheffield, Sheffield S3 7RH, United Kingdom}
\author{S.~L.~Olsen}
\affiliation{Center for Underground Physics, Institute for Basic Science (IBS), Daejeon 34126, Republic of Korea}
\author{B.~J.~Park}
\affiliation{IBS School, University of Science and Technology (UST), Daejeon 34113, Republic of Korea}
\affiliation{Center for Underground Physics, Institute for Basic Science (IBS), Daejeon 34126, Republic of Korea}
\author{H.~K.~Park}
\affiliation{Department of Accelerator Science, Korea University, Sejong 30019, Republic of Korea}
\author{H.~S.~Park}
\affiliation{Korea Research Institute of Standards and Science, Daejeon 34113, Republic of Korea}
\author{J.~C.~Park}
\affiliation{Department of Physics and IQS, Chungnam National University, Daejeon 34134, Republic of Korea}
\author{K.~S.~Park}
\affiliation{Center for Underground Physics, Institute for Basic Science (IBS), Daejeon 34126, Republic of Korea}
\author{S.~D.~Park}
\affiliation{Department of Physics, Kyungpook National University, Daegu 41566, Republic of Korea}
\author{R.~L.~C.~Pitta}
\affiliation{Physics Institute, University of S\~{a}o Paulo, 05508-090, S\~{a}o Paulo, Brazil}
\author{H.~Prihtiadi}
\affiliation{Department of Physics, Universitas Negeri Malang, Malang 65145, Indonesia}
\author{S.~J.~Ra}
\affiliation{Center for Underground Physics, Institute for Basic Science (IBS), Daejeon 34126, Republic of Korea}
\author{C.~Rott}
\affiliation{Department of Physics, Sungkyunkwan University, Suwon 16419, Republic of Korea}
\affiliation{Department of Physics and Astronomy, University of Utah, Salt Lake City, UT 84112, USA}
\author{K.~A.~Shin}
\affiliation{Center for Underground Physics, Institute for Basic Science (IBS), Daejeon 34126, Republic of Korea}
\author{D.~F.~F.~S. Cavalcante}
\affiliation{Physics Institute, University of S\~{a}o Paulo, 05508-090, S\~{a}o Paulo, Brazil}
\author{M.~K.~Son}
\affiliation{Department of Physics and IQS, Chungnam National University, Daejeon 34134, Republic of Korea}
\author{N.~J.~C.~Spooner}
\affiliation{Department of Physics and Astronomy, University of Sheffield, Sheffield S3 7RH, United Kingdom}
\author{L.~T.~Truc}
\affiliation{Department of Physics, Kyungpook National University, Daegu 41566, Republic of Korea}
\author{L.~Yang}
\affiliation{Department of Physics, University of California San Diego, La Jolla, CA 92093, USA}
\collaboration{COSINE-100 Collaboration}


\date{\today}

\begin{abstract}
We report limits on WIMP dark matter derived from three years of data collected by the COSINE-100 experiment with NaI(Tl) crystals, achieving an improved energy threshold of 0.7~keV. This lowered threshold enhances sensitivity in the sub-GeV mass range, extending the reach for direct detection of low-mass dark matter. Although no excess of WIMP-like events was observed, the increased sensitivity enabled a model-insensitive comparison between the expected WIMP signal rate--based on mass limits from our data--and DAMA’s reported modulation amplitude. Our findings strongly disfavor the DAMA signal as originating from WIMP interactions, fully excluding DAMA/LIBRA 3$\sigma$ allowed regions and providing enhanced WIMP mass limits by an order of magnitude in the spin-independent model compared to previous results.
In the spin-dependent model, cross-section upper limits were obtained in the mass range [0.1--5.0]~GeV/c$^2$, with additional sensitivity to sub-GeV WIMPs through the inclusion of the Migdal effect. These results represent substantial progress in low-mass dark matter exploration and reinforce constraints on the longstanding DAMA claim.
\end{abstract}

%\keywords{}

\maketitle


%% Introduction
Dark matter, a nonluminous form of matter, is believed to account for approximately 27\% of the total mass-energy of the Universe, yet its exact nature remains one of the biggest mysteries in physics. While its presence is inferred through gravitational effects on cosmic structures, direct detection of dark matter particles has proven challenging~\cite{Lee:1977ua,Planck:2018pxb}. Among the leading candidates for dark matter are weakly interacting massive particles (WIMPs), which are hypothesized to interact with ordinary matter via the weak interaction~\cite{Jungman:1995df}. 
Despite decades of extensive searches, no conclusive evidence for WIMPs has been observed, with the notable exception of the controversial results from the DAMA/LIBRA experiment, which claims evidence for dark matter interactions using NaI(Tl) crystals~\cite{Bernabei:2018jrt,Bernabei:2022xgg}. However, this result has yet to be corroborated by other direct detection experiments, which have not observed similar signals~\cite{PhysRevD.98.030001}. 
To directly test the DAMA/LIBRA claim, COSINE-100 conducts a dark matter search using the same NaI(Tl) target. Previous publications detail the data acquisition system (DAQ)~\cite{COSINE-100:2018rxe}, noise rejection down to 0.7\,keV~\cite{yu2024lowering}, and background modeling with Geant4~\cite{yu2024improved}. Physics analyses include a model-dependent exclusion of the DAMA/LIBRA 3$\sigma$--allowed region using 1.7 years of data~\cite{COSINE-100:2021xqn}, and a model-independent modulation search based on 6.4 years of data, which shows a discrepancy with DAMA/LIBRA signal at more than 3$\sigma$~\cite{carlin2024cosine}. Additionally, Ref.~\cite{adhikari2023induced} shows that a time-dependent background can produce a modulation similar in amplitude but opposite in phase to that of DAMA/LIBRA when using their analysis method.
Similarly, the ANAIS experiment, also employing NaI(Tl) crystals, reported modulation results that do not support DAMA’s annual modulation signal~\cite{amare2021annual,coarasa2024anais}.
These findings highlight the importance of further increasing sensitivity to low-mass dark matter by lowering the energy threshold and expanding the search to lower WIMP mass ranges~\cite{ibe2018migdal,amole2019dark,angloher2022testing,arora2024search}.
In this letter, we report WIMP mass limits based on a reduced energy threshold of 0.7 keV, advancing constraints on low-mass dark matter and providing a rigorous test of the longstanding DAMA findings.

%% Experiment
COSINE-100\cite{Adhikari:2017esn} is located at the YangYang underground laboratory (Y2L) in South Korea, beneath 700\,m of rock (equivalent to 1890 meters of water)~\cite{LEE2006201,Kim:2012rza}. The experiment consists of eight NaI(Tl) detectors with a total mass of 106~kg. Each crystal is coupled to two 3-inch Hamamatsu R12669SEL photomultiplier tubes (PMTs) via 12~mm quartz light guides. The crystal is wrapped in 10 layers of 250~$\mu$m PTFE reflectors and enclosed in 1.5~mm copper case.
Due to high background levels and low light yields from three crystals, this analysis uses data from only five crystals, with an effective mass of 61.3~kg, labeled C2, C3, C4, C6, and C7~\cite{COSINE-100:2021mrq}. The crystals are immersed in 2200~L of linear alkylbenzene (LAB)-based liquid scintillator (LS) as an active veto detector~\cite{Jengsu:2017,Adhikari:2020asl}. The LS is housed in a 1~cm thick acrylic box, surrounded by 3~cm of copper and 20~cm of lead to shield against external radiation. Plastic scintillators on the outermost layer are used to veto muon-induced events~\cite{COSINE-100:2017jinst,COSINE-100:MuonMod}.

The COSINE-100 experiment operated from September 22, 2016, to March 14, 2023.
This analysis uses data collected up to November 21, 2019, corresponding to a live-time of 2.86 years. The detector, including its energy scale, was verified to be stable throughout the data-taking period~\cite{adhikari2022three}. A validated background model is available above the 0.7\,keV threshold, motivating the use of this dataset.
Events were collected by two PMTs attached to each crystal, with the photon signals amplified in a preamplifier and digitized by an analog-to-digital converter (ADC) at a sampling rate of 500 Megasample per sec. The ADC, connected to the trigger control board (TCB), recorded events satisfying specific trigger conditions, capturing waveforms over an 8~$\mu$s window starting from 2.37~$\mu$s before the trigger time~\cite{COSINE-100:2018rxe}.

%% Analysis
PMT-induced noise is particularly prominent in the low-energy region below 2~keV~\cite{COSINE-100:2020wrv}, where the WIMP is expected to deposit energy in the region of interest (ROI) below 6~keV.
Although the noise is categorized as Type-1 and Type-2, its origin remains unclear. However, it exhibits distinct pulse shapes compared to scintillation signals from electron and nuclear recoils in NaI(Tl), especially in timing characteristics. These differences are captured using pulse-shape discrimination (PSD) parameters in both time and frequency domains. To enhance noise rejection, a multilayer perceptron (MLP) was applied based on these features~\cite{yu2024lowering}.

The MLP was trained using scintillation-rich samples from $\gamma$-rays at 511~keV or 1274~keV from a $^{22}$Na source, tagged by a coincidence condition with high-energy events in neighboring crystals or the liquid scintillator.
A subset of single-hit events from physics data, dominated by PMT-induced noise, was used as the noise sample. 
Using PSD parameters, the MLP could effectively classify events within the WIMP ROI, achieving stringent selection criteria with less than 1\% noise contamination and approximately 20\% scintillation detection efficiency down to 8 photoelectrons, equivalent to an energy threshold of 0.7~keV~\cite{yu2024lowering}.

\begin{figure}  
    \centerline{\includegraphics[width=0.95\linewidth]{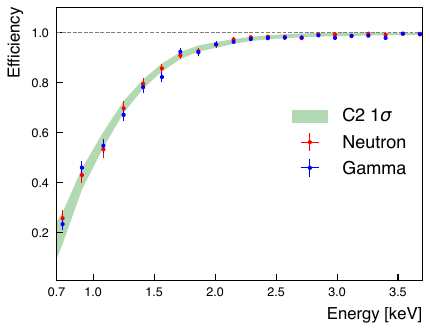}}
    \caption{A comparison of MLP selection efficiency down to 0.7 keV between the 1$\sigma$ band of COSINE-100 C2 and the electron-recoil and nuclear-recoil selection efficiencies from the NaI(Tl) crystal sample.}
\label{fig:ernr}
\end{figure}

Although the MLP was trained on electron-recoil events, it is not expected to distinguish them from nuclear recoils, which exhibit similar or shorter mean-times below 2\,keV~\cite{lee2015pulse}. In contrast, PMT-induced noise shows longer mean-times in this energy range~\cite{COSINE-100:2020wrv}. This was validated by comparing selection efficiencies using $^{137}$Cs and 2.43~MeV neutron beam data~\cite{NaIQF}.
These measurements were conducted on a sample crystal made from the same ingot as the COSINE-100 crystals. The selection efficiency was found consistent between electron and nuclear recoils, as shown in Fig.~\ref{fig:ernr}. The efficiency comparison was conducted with the selection criteria for the sample crystal adjusted to achieve the same selection efficiency as C2.

% Background understanding
With scintillation events dominating the remaining data, quantitatively understanding these events is essential to identify any nuclear recoil events from WIMP-nucleon interactions. In the COSINE-100 detector, scintillation events arise largely from background radioactive isotopes and are categorized as either single-hit or multiple-hit, depending on their coincidence with signals in the liquid scintillator (LS) or another crystal.
Removing multiple-hit events significantly reduces the background, as WIMP-induced signals are expected only in single-hit events.
In particular, the prominent $^{40}$K background in the WIMP search region, with its 3.2~keV X-ray emission, can be tagged with 65–75\% efficiency using LS veto tagging~\cite{Adhikari:2020asl}. Both single-hit and multiple-hit events are used to model the background comprehensively, as multiple-hit events offer valuable insights into the background within single-hit events. 

Background modeling is performed by fitting the data to spectra generated from Geant4~\cite{geant4} simulations of background components~\cite{yu2024improved}. The fit covers the [6--4000]~keV range for single-hit events and the [0.7--4000]~keV range for multiple-hit events, with the [0.7--6]~keV region of the single-hit spectrum extrapolated from higher energy fit results to avoid bias in WIMP signal interpretation. Background components include internal sources within the NaI(Tl) crystal and external sources from detector materials such as PMTs, copper, LS, acrylic, and steel structures. Radioisotopes from the $^{238}$U, $^{232}$Th, $^{40}$K, and $^{235}$U decay chains are considered across these background sources. 
Scintillators like NaI(Tl) crystals are known to exhibit a non-linear relationship between the energy of incident radiation and light output. This non-proportional energy response of the COSINE-100 NaI(Tl) crystals~\cite{COSINEnPR} is applied in developing the energy calibration curves, optimized for the extended threshold region by utilizing the 0.87 keV X-ray peak from $^{22}$Na decay. The resulting calibration curves, along with energy resolution and event selection efficiency for each crystal, are incorporated into the simulations.
The fractional contributions of the background sources are determined by the fit, with some constrained by the measurements obtained from the previous analyses~\cite{Adhikari:2017esn,COSINE-100:2019rvp,COSINE-100:alpha}.
Background studies identify the dominant sources within the WIMP search region as internal $^{40}$K and $^{210}$Pb contamination in the crystal, cosmogenic $^{3}$H, and surface $^{210}$Pb on both the crystal and surrounding PTFE reflector. The activity of $^{3}$H is well understood from cosmogenic production rate calculations~\cite{COSINE-100:2019rvp}, and the contribution from $^{40}$K is characterized by its distinctive 3.2~keV X-ray emission observed in both single- and multiple-hit events~\cite{Adhikari:2020asl}.
The internal $^{210}$Pb activity in the crystal is monitored through the 5.3~MeV $\alpha$ decay measurement from $^{210}$Po~\cite{COSINE-100:alpha}. 
Surface $^{210}$Pb contributions from the crystal and PTFE reflector were assessed by modeling depth profiles from $^{222}$Rn contamination tests~\cite{Yu:2021depth} and analyzing the COSINE-100 alpha background~\cite{COSINE-100:alpha}. These profiles were then applied to simulate the surface background spectrum.

\begin{figure*}[ht]
    \centerline{\includegraphics[width=0.95\linewidth]{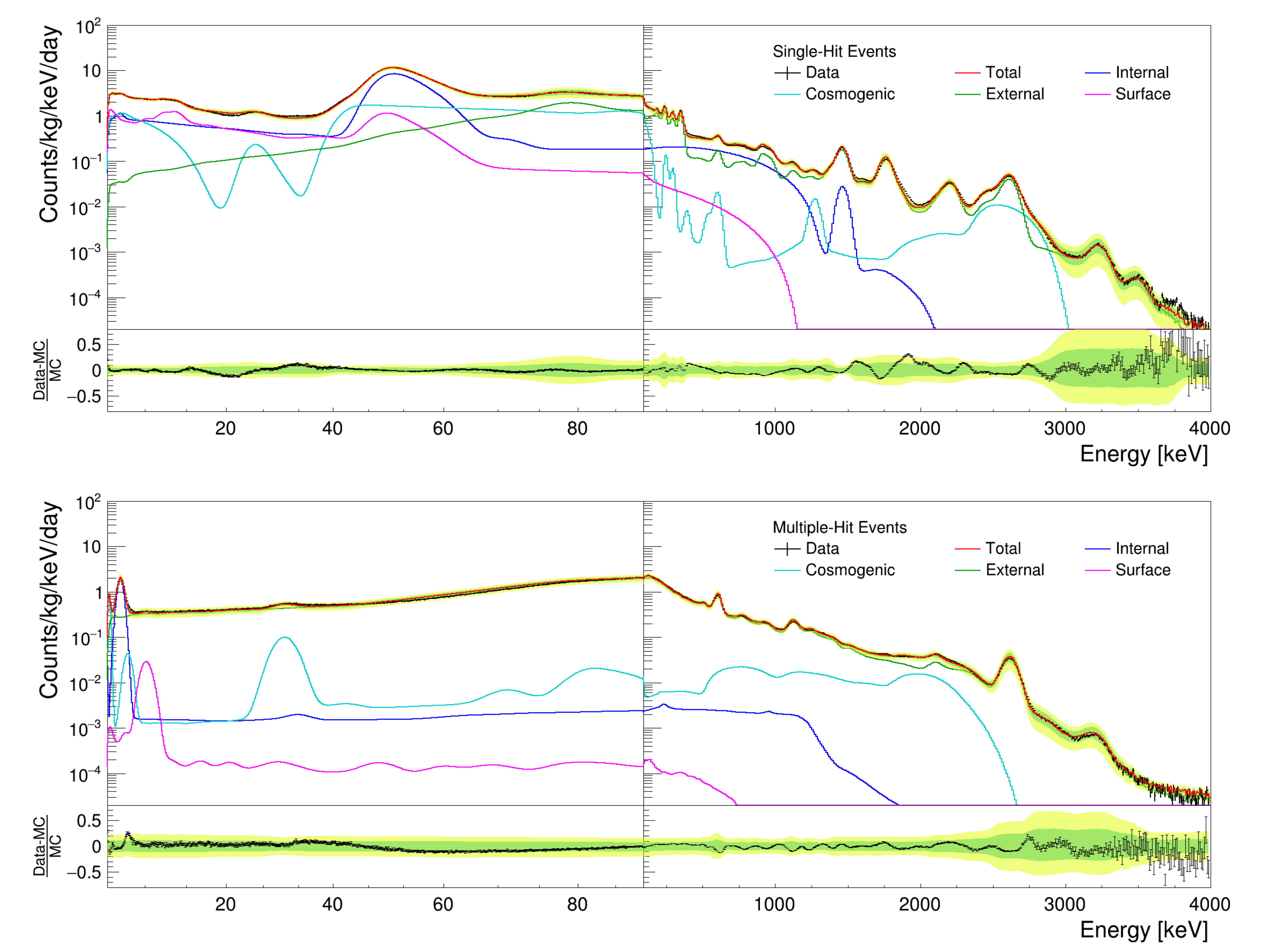}}
    \caption{Results of background modeling for (top) single- and (bottom) multiple-hit events summed over five crystals, shown with the 1 and 2 $\sigma$ systematic uncertainty band.}
\label{fig:systmodel}
\end{figure*}

% Systematics
The background modeling results for single- and multiple-hit events are shown in Fig.~\ref{fig:systmodel}. Differences between data and model across the full energy range of [0.7--4000]~keV remain within the 2\,$\sigma$ systematic uncertainty bands. To validate the fit results, potential sources of systematic uncertainty in the background spectrum were investigated, including: (1) activity uncertainties of background radioisotopes, (2) event selection efficiency, (3) energy-dependent resolution, (4) calibration, (5) surface $^{210}$Pb, and (6) PMT background position.  
Uncertainties in radioisotope activity are derived from fit errors in background modeling, with systematic variations arising from measurement uncertainties in background source activities.
For event selection, statistical uncertainties in MLP signal selection are included, along with efficiency differences between nuclear and electron recoils. Calibration uncertainties account for the non-proportional energy response model, while energy-dependent resolution uncertainties are derived from the fit errors in the resolution function, as detailed in Ref.~\cite{COSINEnPR}.
The spectrum of surface $^{210}$Pb is depth-dependent, and the profile from Ref.~\cite{Yu:2021depth} is used to account for depth-related uncertainties.
Additionally, while the PMTs attached to the crystals are contaminated with radioactive sources, the precise location of this contamination within the PMTs remains unclear. The energy deposition spectrum from these contaminants varies with distance from the crystal, so the spectral uncertainty due to PMT position (from the window to the base) is also considered.

%% Results
To estimate the contribution of nuclear recoil events from dark matter interactions in the data, the WIMP energy spectrum was simulated using the publicly available \textsc{DMDD} package~\cite{gluscevic2015dmdd}.
The nuclear recoil rate, when recoil energy is $E_{nr}$, is calculated using the following equation:
\begin{equation}
\frac{dR}{dE_{nr}} = 
\frac{\rho_{\chi}}{2m_{\chi}\mu^2}\sigma(q)
\int_{v_{min}}^\infty d^3v 
\frac{f(\mathbf{v},t)}{v},
\end{equation}
where $m_{\chi}$ is the dark matter mass, $\mu$ is the reduced mass of the dark matter particle and the target nucleus, $v$ is the dark matter velocity, and $\sigma(q)$ is the WIMP-nucleus scattering cross section as a function of the momentum transfer $q$~\cite{billard2022direct}. 
Here, we assume a velocity-independent scattering cross section, as in standard spin-(in)dependent interactions. The local dark matter density is set to 0.3\,GeV/$c^2$, and time-dependent velocity distribution $f(\mathbf{v},t)$ follows the standard halo model (SHM).
%We assume the standard halo model (SHM), setting the local dark matter density to 0.3\,GeV/$c^2$, and time dependent velocity distribution $f(\mathbf{v},t)$. The $\sigma(q)$ is expressed as
\begin{equation}
    \sigma(q) = \sigma_{SI}F^2_{SI}(q) + \sigma_{SD}F^2_{SD}(q),
\end{equation}
while $F(q)$ denotes the nucleus form factor. 
The cross-section for Spin-Independent (SI) interaction is defined as 
\begin{equation}
    \sigma_{SI} = \frac{4}{\pi}\mu^2[Zf_p + (A-Z)f_n]^2,
\end{equation}
when $f_p$ and $f_n$ are coupling constants to protons and neutrons, respectively, and A and Z are the mass and proton numbers of the nucleus. 
Similarly, the cross-section for Spin-Dependent (SD) interaction is defined as
\begin{equation}
    \sigma_{SD} = \frac{32}{\pi}G^2_{F}\mu^2(\frac{J+1}{J})[\langle S_p \rangle a_p + \langle S_n \rangle a_n]^2,
\end{equation}
when $G_{F}$ is the Fermi coupling constant, $J$ is the ground state angular momentum of the nucleus, $\langle S_{p} \rangle$ and $\langle S_{n} \rangle$are the average proton and neutron spin contributions, and $a_{p}$ and $a_{n}$ are the axial four-fermion WIMP-proton and WIMP-neutron couplings. 
In this study, the SI model for the isospin-conserving case, where $f_p \sim f_n$, has been considered. It is noteworthy that the NaI(Tl) target can be employed for both SI and SD interactions due to the proton-odd compositions of $^{23}$Na and $^{127}$I, which results in a relatively high value for $\langle S_p \rangle$. 
Additionally, the simulation of the Migdal effect~\cite{migdal1941ionization,ibe2018migdal,MigdalXenon}, an ionization process induced by the sudden momentum change of the target nucleus, has also been performed~\cite{PhysRevD.105.042006}.
%The simulated nuclear recoil spectrum was converted to electron-equivalent energy by applying the quenching factor (QF) measured in Ref.~\cite{lee2024measurements}, covering nuclear recoil energies down to 3.8\,keV for Na and 6.1\,keV for I recoils. These correspond to electron-equivalent energies of 0.43\,keV and 0.30\,keV, respectively, accounting for the non-proportional energy response, encompassing the 0.7\,keV energy threshold.
%
%The simulated nuclear recoil spectrum was converted to electron-equivalent energy using newly measured quenching factors (QF) from Ref.~\cite{lee2024measurements}. These measurements, extending down to 3.8\,keVnr for Na ($Q_{Na}$=11.2\%) and 6.1\,keVnr for I ($Q_{I}$=4.9\%), correspond to 0.43 and 0.30\,keVee, respectively--well below the 0.7\,keV analysis threshold. Unlike DAMA/LIBRA, which used fixed values of 30\% for Na and 9\% for I over the entire energy range, COSINE-100 applies energy-dependent QF, accounting for the non-proportional scintillation response.
The simulated nuclear recoil spectrum was converted to electron-equivalent energy using newly measured quenching factors (QF) from Ref.~\cite{lee2024measurements}. These measurements, extending down to 3.8\,keVnr for Na ($Q_{Na}$=11.2\%) and 6.1\,keVnr for I ($Q_{I}$=4.9\%), correspond to 0.43 and 0.30\,keVee, respectively--well below the 0.7\,keV analysis threshold.
Unlike DAMA/LIBRA, which used fixed values of 30\% for Na and 9\% for I over the entire energy range, COSINE-100 applies energy-dependent QF, accounting for the non-proportional scintillation response. %\textcolor{red}{
Because different QF assumptions are used, the COSINE-100 upper limits shift upward relative to the DAMA QF case, resulting in a more conservative exclusion. A direct comparison under identical QF assumptions was reported in our previous 1.7-year analysis~\cite{COSINE-100:2021xqn}.
%}

\begin{figure}[t]
    \centerline{\includegraphics[width=1.0\linewidth]{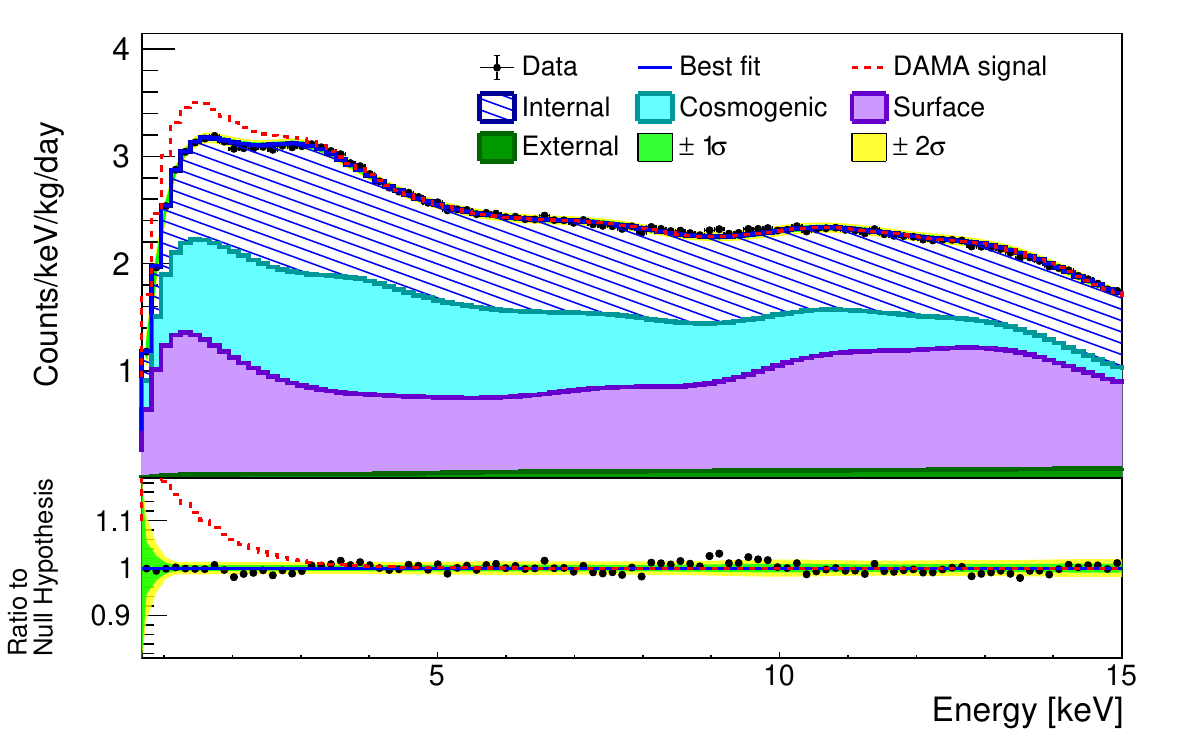}}
    \caption{Fit example of WIMP presence test on COSINE-100 3-year data, considering SI model and $m_{\chi}$=11.5\,GeV/$c^2$. The expected spectrum based on DAMA/LIBRA's observation($\sigma_{\text{SI}}=1.5\times10^{-4}$~pb, without Migdal) is shown as a red dashed line.}
    \label{fig:fitexample}
\end{figure}

The probability of a WIMP signal in the data was evaluated using a likelihood function based on Bayes' theorem, with the background components, systematics, and WIMP signal as input parameters~\cite{COSINE-100:2021xqn}. The posterior distribution of the WIMP-nucleon cross section was derived by the multi-variable fit procedure based on the Markov Chain Monte Carlo sampling method~\cite{gilks1995markov,gamerman2006markov}, implemented via Metropolis-Hastings algorithm~\cite{metropolis1953equation, hastings1970monte}. The fit was performed on the single-hit energy spectrum in the 0.7 to 15~keV range for WIMP signals from both SI and SD interaction models across several WIMP masses. %\textcolor{red}{The impact of the fit range was evaluated and found to be negligible.}
The impact of the fit range was evaluated and found to be negligible.
The means and uncertainties of the fractional activities for the background components, determined from modeling, were used as Gaussian priors, and systematic uncertainties were treated as nuisance parameters with Gaussian priors. The background and systematic parameters were allowed to vary independently in each crystal channel, while the WIMP signal strength was constrained simultaneously across all crystals. The resulting posterior distribution of the signal strength was found to be consistent with a no-signal scenario in both SI and SD channels. 
The fit spectrum for an $m_{\chi}$=11.5\,GeV/$c^2$ WIMP at a 90\% confidence level (C.L.) is shown as a solid blue line in Fig.~\ref{fig:fitexample} (upper pannel) with the summed spectrum from the five crystals represented by black cross markers. The dashed red line represents the simulated WIMP signal combined with the background, assuming a WIMP-nucleon cross-section of 1.5$\times10^{-4}$~pb for a $m_{\chi}$=11.5\,GeV/$c^2$ WIMP in the SI model, which corresponds to the lower bound of the DAMA/LIBRA 3$\sigma$ contour~\cite{savage2009compatibility} using DAMA's QF~\cite{Bernabei:1996vj} for WIMP-sodium interactions. 
The lower panel shows the ratio of the data to the best fit with the 1$\sigma$ and 2$\sigma$ bands of systematic uncertainty. The ratio of the DAMA/LIBRA signal to our best fit is also shown as a dashed red line.
\begin{figure*}[ht]
\centering
\begin{tabular}{cc}
\includegraphics[width=0.49\linewidth]{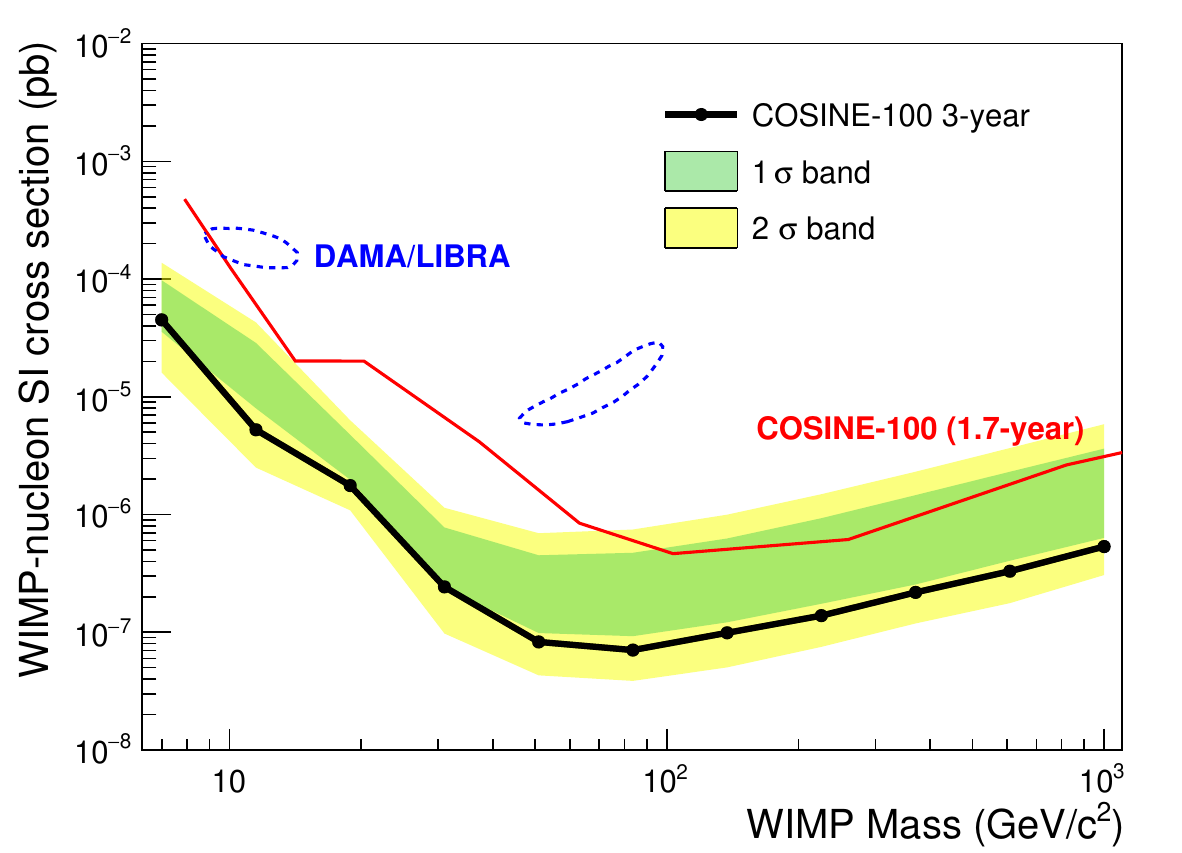} &
\includegraphics[width=0.49\linewidth]{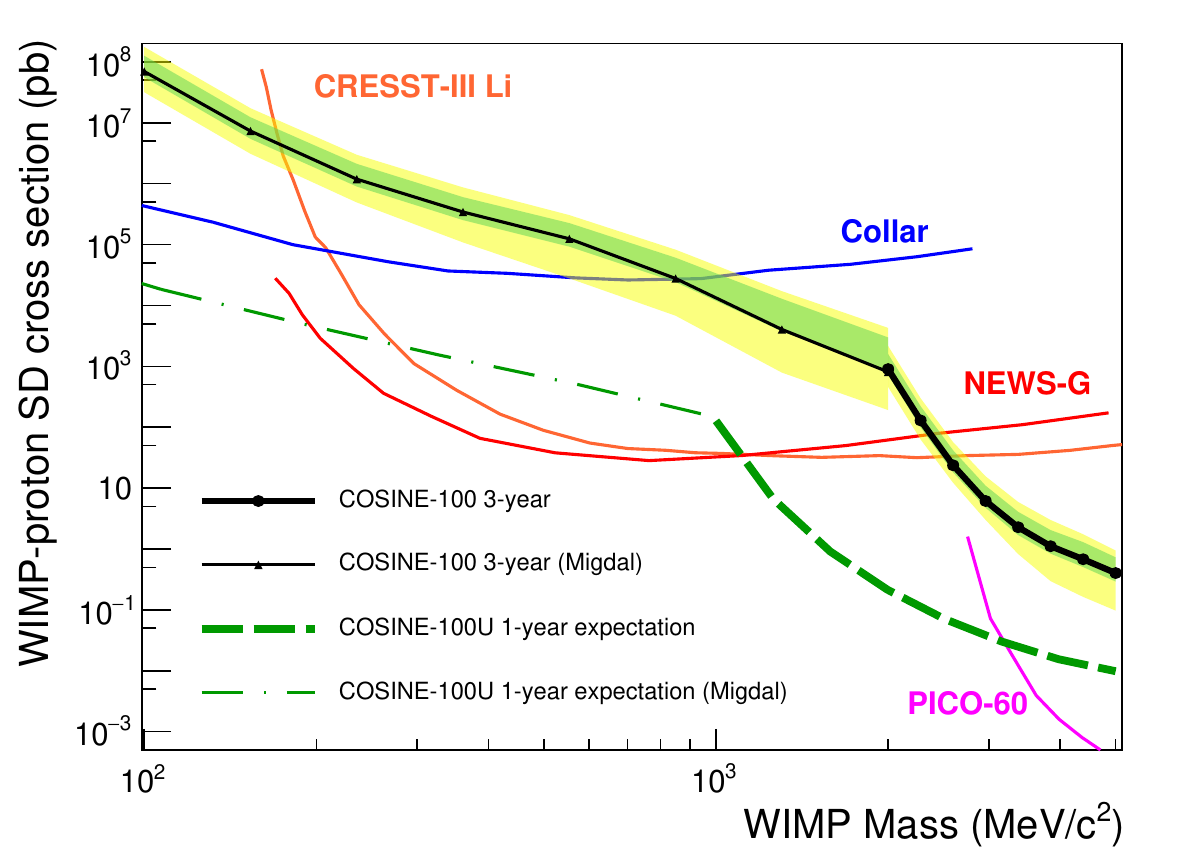} \\
(a) & (b) \\
\end{tabular}
%\caption{(a) WIMP-nucleon SI interaction (b) WIMP-proton SD interaction.}
\caption{%\textcolor{red}{
The COSINE-100 3-year 90\% C.L. upper limits on (a) SI and (b) SD WIMP-nucleon cross sections. In (a), the DAMA/LIBRA 3$\sigma$ region is overlaid for comparison. In (b), leading experimental constraints are shown, and the COSINE-100 SD limits, including the Migdal effect, extend sensitivity to the sub-GeV masses. Furthermore, more stringent limits are expected from COSINE-100U.}
%}
\label{fig:upper_limit}
\end{figure*}
%

%The excess events from the DAMA/LIBRA signal were evaluated by comparing DAMA's observed modulation amplitude ($A_\text{DAMA}$) with the exclusion limit on the total expected signal rate ($R_\text{COSINE}$) from this analysis as a model-independent comparison, following the approach suggested by the COSINUS experiment~\cite{cosinus_mod}.
%\textcolor{red}{
The excess events in the DAMA/LIBRA signal were evaluated by directly comparing the observed modulation amplitude ($A_\text{DAMA}$) with the total event rate limit ($R_\text{COSINE}$) from this analysis. While the calculation of $R_\text{COSINE}$ assumes a WIMP mass of 11.5\,GeV/$c^2$, the comparison itself is considered model-insensitive, as it contrasts two measured quantities with minimal reliance on specific model assumptions. This approach follows the method proposed by the COSINUS experiment~\cite{cosinus_mod}.
%}
The values of $R_\text{COSINE}$ and $A_\text{DAMA}$ are compared across the energy ranges of [1--3], [1--6], and [2--6]\,keV. 
The $R_\text{COSINE}$ values for a $m_{\chi}$=11.5\,GeV/$c^2$ WIMP at the 90\% C.L. were calculated to be (0.0048, 0.0020, 0.0006)  in three energy ranges, while DAMA's modulation amplitudes are (0.019~$\pm$~0.002, 0.010~$\pm$~0.001, 0.0096~$\pm$~0.0007). 
Additionally, we performed this comparison across the entire WIMP mass range from 7 to 1000\,GeV/$c^2$, observing consistent rates, with the maximum rate being (0.0067,  0.0031, 0.0009) at $m_{\chi}$=18.8\,GeV/$c^2$.
The expected WIMP signal rates are generally about 10\% of DAMA’s modulation amplitude and reach about 30\% at maximum, making them significantly lower than the $A_\text{DAMA}$ values across all three energy ranges.
This suggests that the annual modulation signals observed by DAMA are unlikely to arise from WIMP interactions.

%To calculate the expected 90\% C.L. upper limit on the WIMP-nucleon SI interaction cross-section as a function of WIMP mass, 1000 pseudo-experiments were generated under the null hypothesis. The exclusion limits were then obtained by fitting the simulated WIMP signal and background model to the pseudo-experiment data, resulting in a distribution of 1000 upper limits at the 90\% C.L. The 1$\sigma$ and 2$\sigma$ bands, derived from this distribution, are shown in green and yellow areas in Fig.~\ref{fig:upper_limit}. 
To calculate the expected 90\% C.L. upper limit on the WIMP–nucleon SI interaction cross section as a function of WIMP mass, %\textcolor{red}{
1000 pseudo-experiments were generated under the null hypothesis by Poisson sampling from the modeled background spectrum. The systematic effects were propagated by drawing the associated components within their quoted uncertainties to obtain the expected rates, from which event counts were Poisson-sampled. These systematic uncertainties were also included in the fit of the simulated WIMP signal and background model to the pseudo-experiment data as nuisance parameters with Gaussian constraints~\cite{COSINE-100:2021xqn}. The resulting distribution of 90\% C.L. upper limits defines the 1$\sigma$ and 2$\sigma$ expected sensitivity bands, shown as the green and yellow regions in Fig.~\ref{fig:upper_limit}.
%}
The observed 90\% C.L. upper limits from 3 years of COSINE-100 data are represented by filled black circles. 
%The results for the SI model, covering WIMP masses ranging from 7 to 1000\,GeV/$c^2$, is presented Fig.~\ref{fig:upper_limit}(a). Here, the previous COSINE-100 data is shown as a red solid line, while the DAMA/LIBRA 3$\sigma$ allowed region, fully excluded in this analysis, is depicted as a dashed blue line.
Figure~\ref{fig:upper_limit}(a) presents the results for the SI model, covering WIMP masses ranging from 7 to 1000\,GeV/$c^2$. The previous COSINE-100 data is shown as a red solid line, while the DAMA/LIBRA 3$\sigma$ allowed region, fully excluded in this analysis, is depicted as a dashed blue line. 
The previous 1.7 years' results excluded all the DAMA/LIBRA 3$\sigma$ regions when different QFs were applied~\cite{COSINE-100:2021xqn}.
%textcolor{red}{The previous 1.7-year results excluded all DAMA/LIBRA 3$\sigma$ regions under both the same and different QF assumptions~\cite{COSINE-100:2021xqn}, while the estimated DAMA signal cross section is lower in the DAMA QF scenario.}
This analysis achieves WIMP mass limits that are improved by an order of magnitude over the previous result. 
The SD model results, shown in Fig.~\ref{fig:upper_limit}(b), focus on WIMP masses ranging from 0.1 to 5\,GeV/$c^2$, providing competitive cross-section limits around 2.5\,GeV/$c^2$ compared to other experimental searches~\cite{collar2018search,amole2019dark,angloher2022testing}. Particularly, the inclusion of the Migdal effect significantly enhances sensitivity to the low-mass WIMPs below 1~GeV/c² in the SD model. The sensitivity of the COSINE-100U experiment, an upgraded version of COSINE-100 with a $\sim40\%$ enhancement in light yield~\cite{lee2024cosine}, is expected to probe unexplored parameter spaces for WIMP masses, potentially reaching extremely low-mass regions down to 20\,MeV/$c^2$ when considering the Migdal effect.
While theoretical uncertainties to the Migdal effect are not evaluated in this work, understanding their impact will be important for future low-mass dark matter searches.

\begin{acknowledgments}
We thank the Korea Hydro and Nuclear Power (KHNP) Company for providing underground laboratory space at Yangyang. This work is supported by the Institute for Basic Science (IBS), South Korea; UIUC, the Alfred P. Sloan Foundation, NSF, WIPAC, the Wisconsin Alumni Research Foundation, Yale University, DOE/NNSA, STFC, CNPq, and FAPESP, Brazil.
\end{acknowledgments}

\bibliographystyle{apsrev4-1}
\bibliography{newbib} % Assuming your .bib file is named "references.bib"

\end{document}